# The comparative analysis of channels of three α-particles and $^{12}$C nuclei production in $^{16}$Op-collisions at 3.25 A GeV/c

M.L. Allaberdin, M.A. Belov[1], K.G. Gulamov, V.D. Lipin, V.V. Lugovoi, S.L. Lutpullaev,
K. Olimov, Kh.K. Olimov, A.A. Yuldashev, B.S. Yuldashev[1]

Physical-technical institute of SPA "Physics—Sun" of Uzbek Academy of Sciences,
G.Mavlyanova str., 2B, Tashkent, 700084, Uzbekistan

## 1. Introduction

Experimental data on the yield of multinucleon nuclei in interactions of hadrons and nuclei with nuclei can have valuable information about fragmentation mechanisms and nuclei structure.

It is known, that the light nuclei with mass numbers A≤20, multiple 4, have an α-cluster structure and many characteristics of them are satisfactorily described by models [1], taking into account this feature. At the investigation of processes of fragmentation of relativistic oxygen nuclei in interaction with a proton [2-7] some features of this phenomenon have been determined. In particularly also has been showed, that among multicharged fragments twocharged ones have the most probable yield and more than 80 % of which are $^{4}$He nuclei, i.e. α-particles [3]. The decay of relativistic oxygen nuclei into multicharged fragments with the conservation in them of a charge of initial nucleus [5] is realised only by topologies with the even charge (224), (2222) and (26), where charges of nuclei-fragments specified in brackets. In the same time channels (44), (35) and (233) are absent. Qualitatively it can be explained, that the breakdown of a residual weakly excited nucleus with the charge Z=8 most probably occurs on α-particles as in a case (2222) (a decay on the channel (44) as a result of breakdown of an unstable nucleus $^{8}$Be also leads to topology (2222)), or on twocharged fragment and carbon nucleus. The decay of oxygen $^{16}$O nuclei into two and more multicharged fragments at conservation in them both the charge and the total number of nucleons of an initial nucleus [6] only two channels is observed: four α-particles or $^{12}$C and $^{4}$He nuclei, i.e. the decay of $^{16}$O is realised with the production of even-even nuclei.

---

[1] Institute of Nuclear Physics, Uzbek Academy of Sciences, pos. Ulugbek, Tashkent, 702132 Republic of Uzbekistan.

It is clear from the above mentioned experimental results that the structure of an initial nucleus is shown basically at peripheral collisions. In this connection it becomes interesting to carry out the comparative analysis of characteristics of an oxygen nucleus fragments in channels, concerning small energy transfers, i.e. at low excitation energies of an oxygen nucleus in interactions with a proton.

The present work is devoted to the comparative analysis of the characteristics of reactions

$$^{16}O + p \rightarrow 3\alpha + X, \qquad (1)$$

$$^{16}O + p \rightarrow {}^{12}C + X. \qquad (2)$$

## 2. Experimental data and their discussion

The experimental material is received with the help of stereosnapshots at one-meter hydrogen bubble chamber (HBC) of LHE, JINR, irradiated by oxygen−16 nuclei with a momentum 3.25 A GeV/c and is based on the analysis of 11098 measured $^{16}$Op-events. Methodical questions on the processing of stereophotos from HBC are described in works [3,4]. Let us note, that experimental conditions allow to register all secondary charged particles, to identify unequivocally their charges, to measure with high precision a momentum and to define fragments masses.

The selection of fragments on mass was carried out on the measured value of momentum and charge. The fragments with the measured length of a track L > 35 cm were considered; that was necessary for their more reliable selection on mass and analysis of their kinematical characteristics. With the purposes of final fragments identification on mass, following momentum intervals were introduced: (4.75-7.8) GeV/c for $^2$H; P > 7.8 GeV/c for $^3$H; P < 10.8 GeV/c for $^3$He and 10.8 < P < 16.5 GeV/c for $^4$He. Sixcharged fragments with momenta 37 < P < 41 GeV/c were referred to $^{12}$C nuclei. Singlecharged relativistic positive particles with momenta 1.75 < P < 4.75 GeV/c were related to protons - fragments. The separation of recoil protons and $\pi^+$-mesons was carried out in the momenta range P < 1.25 GeV/c. Thus, it was left 411 and 488 events, satisfying to these conditions for reactions (1) and (2) respectively. With the taking into account of events loss on inter-action of α-particles and $^{12}$C at a run length L=35 cm in a chamber working volume finally we get the following values of yield cross section for these reactions: $\sigma_{in}(3\alpha)$=(26.5±1.6) mbn and $\sigma_{in}(^{12}C)$=(27.5±1.8) mbn, which within the limits of experimental errors coincide with each other.

The difference on the threshold energy of reactions (1) and (2) at identity of particles of "X"–type and their kinematics characteristics does not exceed 7.4 MeV. This energy is insufficient for the decay of residual nucleus with $A_f$=12 on others channels, except the reaction (1), including also the cascade decay ($^{12}C^* \rightarrow {}^8Be^* + \alpha$, $^8Be^* \rightarrow 2\alpha$). In this connection, apparently, it is possible to expect, that many characteristics of reactions (1) and (2) will be identical.

In Table 1 average multiplicities of secondary charged particles and fragments, produced in considered reactions, are given. It is obvious, that average multiplicities of secondary charged particles and fragments within the bounds of statistical errors coincide in both reactions.

**Table 1**
Average multiplicities of secondary particles
and fragments in reactions (1) and (2)

| Particle type | Average multiplicity | |
|---|---|---|
| | $^{16}O + p \rightarrow 3\alpha + X$ | $^{16}O + p \rightarrow {}^{12}C + X$ |
| $\pi^-$ | 0.30±0.03 | 0.33±0.03 |
| $\pi^+$ | 0.52±0.04 | 0.49±0.04 |
| recoil $p$ | 0.53±0.03 | 0.56±0.03 |
| $p$-fragments | 1.40±0.06 | 1.43±0.06 |
| $d$ | 0.23±0.02 | 0.23±0.02 |
| $t$ | 0.04±0.01 | 0.04±0.01 |
| $^3He$ | 0.03±0.01 | 0.03±0.01 |

Average values of total and transverse momenta of secondary charged particles and fragments, produced in reactions (1) and (2) are given in Table 2. Average momenta of sec-ondary particles are submitted in the laboratory system and for fragments - in the oxygen nucleus rest system .

It is clear from Table 2, that except for an average momentum of $\pi^-$-mesons, all characteristics of secondary particles and fragments within the limits of statistical errors coincide in both reactions. It is qualitatively possible to explain distinction between average values of total momenta of $\pi^-$-mesons $<P_{\pi^-}>$ in reactions (1) and (2) as follows. At interaction of a primary proton with one of $\alpha$-clusters the residual nucleus with quantum numbers of three $\alpha$-clusters is produced and during that the production of $\pi^-$-meson is possible. If in the system of $^{16}O$ nucleus rest a velocity of $\pi^-$-meson is close to a residual nucleus one, they can interact with larger probability. Thus, $\pi^-$-meson loses a part of a momentum and invariant mass of the residual nucleus increases and that results in its decay. The similar mechanism of a momentum loss for positive particles is impossible because of coulomb repulsive forces between them, therefore in reactions (1) and (2) their average momenta coincides.

**Table 2**
Average values of total and transverse momenta of secondary charged particles and fragments in reactions (1) and (2)

| Particle type | $^{16}O + p \to 3\alpha + X$ | | $^{16}O + p \to {}^{12}C + X$ | |
| --- | --- | --- | --- | --- |
| | $<P>$ | $<P_\perp>$ | $<P>$ | $<P_\perp>$ |
| $\pi^-$ | 493±14 | 185±11 | 633±33 | 211±10 |
| $\pi^+$ | 538±19 | 236±13 | 519±20 | 223±14 |
| recoil $p$ | 679±20 | 394±13 | 683±20 | 390±12 |
| $p$-fragments | 324±10 | 237±8 | 323±10 | 246±9 |
| $d$ | 245±20 | 249±21 | 243±21 | 246±21 |

In reaction (2) an average momentum of $\pi^+$-meson is less, than a $\pi^-$-meson one. It can be explained by restriction of the top border of an identification of a $\pi^+$-meson momentum (see above) on the one hand, and possible contribution of the inelastic recharge process of a proton-target into $\pi^+$-meson and neutron − on the other.

It is necessary to note a similarity (within the limits of statistical errors) of average values of total and transverse momenta of protons-fragments and deuterons irrespective to a type of considered reactions.

Given experimental data confirm the presence of α-cluster structure in an $^{16}$O nucleus. However, it is not clear, whether three α-particles in reaction (1) are the product of the decay of an excited residual nucleus with quantum numbers of three α-particles (that is $^{12}$C*) or each of them was produced as a result of direct multifragment breakdown of an oxygen $^{16}$O nucleus without any production of the excited condition $^{12}$C*. Though both the first, and the second mechanisms are possible. If so, what part of them is the product of the first mechanism and what contribution of the second one? In order to answer these questions we realised the Monte Carlo (MC) modelling of the process of three α-particle production in reaction (1) in the frame of isotropic phase space model. Furthermore it is suggested, in the rest system of excited $^{12}$C* nucleus its decay into three α-particles should be isotropic. Therefore, we have compared experimental data on reaction (1) to results of calculation by Monte Carlo model (see *Appendix*), which takes into account two channels of isotropic breakdowns[2]:

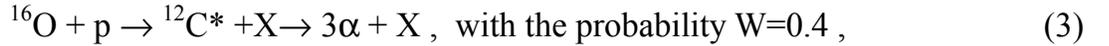
$$^{16}O + p \to {}^{12}C^* + X \to 3\alpha + X, \text{ with the probability } W=0.4, \quad (3)$$

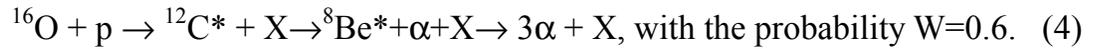
$$^{16}O + p \to {}^{12}C^* + X \to {}^{8}Be^* + \alpha + X \to 3\alpha + X, \text{ with the probability } W=0.6. \quad (4)$$

In the reaction (4) a decay of an unstable $^8$Be* nucleus on two α-particles was generated with the probability W=0.67 for the basic state ($J^P=0^+$) with the excitation energy $\Delta E^*=0.1$ MeV and with the probability W=0.33 for the first excited state ($J^P=2^+$) with $\Delta E^*=3.04$ MeV [8]. In reactions (3) and (4) decays of excited nuclei were generated via the model of isotropic phase space.

Experimental and computed dependences of an average transverse momentum of α-particles $\langle P_T^\alpha \rangle$ on the excitation energy, defined as $\Delta E^* = M(^{12}C^*) - 3M_\alpha$, where $M(^{12}C^*)$ - invariant and $3M_\alpha$ - total mass of three α-particles are submitted in Fig.1. Experimental values of an average transverse momentum $\langle P_T^\alpha \rangle$ depending on $\Delta E^*$ increase linearly in a range of small excitation energy values ($\Delta E^* < 15$ MeV) and, since $\Delta E^* > 15$ MeV their rate of growth appreciably weakens. It, apparently, specifies the fact, that in a range of $\Delta E^* > 15$ MeV a knockout of one α-clusters from weakly bound residual nucleus with three α-clusters occurs, thus not allowing an excited $^{12}C^*$ nucleus to be produced; it results in weak correlations between $\langle P_T^\alpha \rangle$ and $\Delta E^*$.

---
[2] Data on contributions of unstable Be$^8$ nuclei in α-particles production in channels with two, three and four α-particles will be published in one of our subsequent works.

The dependence $\langle P_T^\alpha \rangle$ on $\Delta E^*$ is stronger in theoretical MC calculation, than in experiment, i.e. in the last one the deviation from isotropic decay of the system is observed, that can be connected to simultaneous increase of an average longitudinal momentum of an α-particle. It is confirmed (see Fig.2) by the comparison of the dependence of average absolute longitudinal momenta values of α-particles (in the system of a zero longitudinal momentum of a nucleus $^{12}C^*$ fragment) on $\Delta E^*$ in MC account and experiment. It is interesting to note, that in Fig.1 and 2 at every fixed $\Delta E^*$ the difference $\Delta P_T = \langle P_T \rangle^{MC} - \langle P_T \rangle^{exp}$ between experimental and theoretical values is approximately equal to a difference $\Delta P_L = \langle |P_L| \rangle^{MC} - \langle |P_L| \rangle^{exp}$, taken with an opposite sign, i.e. there is a kinematic compensation: $\Delta P_T = -\Delta P_L$.

It is also obvious from Fig.1 and 2, that at excitation energy values $\Delta E^* < 15$ MeV experimental data within the limits of statistical errors coincide with results of MC calculation. Therefore, if the excitation energy of the system $^{12}C^*$ $\Delta E^* < 15$ MeV (it is ≈51% of the data), the decay $^{12}C^* \to 3\alpha$ can proceed isotropically. However, if $\Delta E^* > 15$ MeV (it is ≈49 % of the data), as it was mentioned above, there is an α-cluster knocking-out process. Data on Fig.3 and 4 specify the presence of such process; on figures the distributions on a difference of azimuthal angles $\Delta\psi$ of α-particles pair for excitation energy $\Delta E^* < 15$ MeV and $\Delta E^* > 15$ MeV respectively are presented. For comparison in Fig.3 data from the work [9], obtained on nuclear photoemulsion, are given, in which the spectrum $\Delta\psi_{\alpha\alpha}$ in reaction

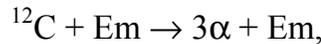
$^{12}C + Em \to 3\alpha + Em,$

is investigated, i.e. in a process of diffraction decay of $^{12}C^*$ nucleus into three α-particles. Thus, a number of combinations from [9] were normalised on a number of combinations of our experiment. It is clear, that within the limits of statistical errors both distributions coincide. It can also serve as an indication on the fact, that at $\Delta E^* < 15$ MeV observable in experiment three α-particles are the product of decay of weakly excited $^{12}C^*$ nucleus. Reasons of a deviation of $\Delta\psi_{\alpha\alpha}$-distribution shape from the isotropic one (a peak at $\Delta\psi_{\alpha\alpha} \leq 18°$ in Fig.3) can be: a cascade decay of $^{12}C^*$ nucleus into three α-particles ($^{12}C^* \to {}^8Be^* + \alpha$, $^8Be^* \to 2\alpha$), a presence of the angular moment at dissociating nucleus and, at last, dynamic mechanisms of interaction in a final state between α-particles (effects of identity). The analysis of distribution of an angle between α-particles pairs $\theta_{\alpha\alpha}$ with $\Delta\psi_{\alpha\alpha} \leq 18°$ at $\Delta E^* < 15$ MeV has shown, that from 144 combinations 133 have $\theta_{\alpha\alpha} < 0.5°$.

The distribution on $\Delta\psi_{\alpha\alpha}$ (Fig.4) at $\Delta E^* > 15$ MeV has two peaks: the first one at $\Delta\psi_{\alpha\alpha} \leq 18°$, and the second - at $\Delta\psi_{\alpha\alpha} = 180°$. The first peak is caused by the mentioned above reasons, and the second one – by knockout of an α-cluster, which transverse momentum in the rest system of $^{12}C^*$ is lager than $^{12}C^*$ transverse momentum in the laboratory system.

Thus, it is possible to conclude, that the reaction (1) is realised in a half part of events as a result of decay of an excited $^{12}C^*$ nucleus, and the rest part of (1) – by the knocking-out of one α-cluster from the weakly bound residual nucleus, containing three α-particles.

The modelling of the decay of an excited system on the model of isotropic phase space is justified at small values of excitation energy and, nevertheless, it allows to take the valuable additional information about the dynamics of investigated process from experimental data.

# Appendix

## 1. Monte Carlo modeling

We suppose three α-particles were produced by decay of one nucleus fragment $^{12}C^*$. The invariant mass distribution of $^{12}C^*$ and distributions of its $P_x$, $P_y$, $P_z$ momentum projections were generated, according to experimental spectra for the system of three α-particles from the reaction (1). In the frame of our Monte Carlo calculations the invariant mass of three α-particles was generated first. Then – momentum projections $P_x$, $P_y$, $P_z$ of vector sum of momentum vectors of three α-particles. In the oxygen nucleus rest frame ($K_0$ frame) the vector sum of momentum of vectors of three α-particles is the momentum vector $\vec{P}_0$ of $^{12}C^*$ nucleus fragment, i.e. $P_0 = \sqrt{P_x^2 + P_y^2 + P_z^2}$. The energy $E_0$ of $^{12}C^*$ nucleus fragment in this system is $E_0 = \sqrt{P_0^2 + M_{3\alpha}^2}$.

## 1.1 The decay : $^{12}C^*$ nucleus fragment → 3α

The process (3) of decay of $^{12}C^*$ nucleus fragment into three α-particles is generated in its rest system. For the transition from $K_0$ (oxygen rest frame) to $^{12}C^*$ nucleus fragment rest frame we turn the $K_0$ frame to obtain the z* axis along to the $^{12}C^*$ nucleus fragment momentum $\vec{P}_0$ and the y* axis along to the vector product $\vec{z}^* \times \vec{z}_0$. Let us define this new system as K* one, and let the K* frame move with $\vec{\beta}_0 = \vec{P}_0/E_0$ velocity. Thereby we obtain the K rest frame of the $^{12}C^*$ nucleus fragment.

The decay (3) of $^{12}C^*$ nucleus fragment into three α-particles is generated according to the isotropic phase space model:

$$d^5W \propto \Phi(M_{12}) \frac{p_3^2}{\varepsilon_3} dp_3 \sin\theta_3 \, d\theta_3 \, d\phi_3 \, \sin\theta_1' \, d\theta_1' \, d\phi_1' \qquad (5)$$

Here $p_3$ and $\varepsilon_3$ are the $\alpha_3$ momentum and energy in the $^{12}C^*$ nucleus fragment rest frame, $M_{12}$ is the invariant $\alpha_1\alpha_2$ mass, $\Phi$ is the phase volume of the $\alpha_1\alpha_2$ system, $\theta_3$ and $\phi_3$ are the production angles of $\alpha_3$ particle in $^{12}C^*$ nucleus fragment rest frame; $\theta_1'$ and $\phi_1'$ are the production angles of $\alpha_1$ particle in the $\alpha_1\alpha_2$ center of mass system (the choice of the $\alpha_1\alpha_2$ center of mass system in the $^{12}C^*$ nucleus fragment rest frame is similar to the choice of the $^{12}C^*$ nucleus fragment rest frame in the $K_0$ system.

After Lorentz transformation of the momenta vector projections of $\alpha_1$, $\alpha_2$, $\alpha_3$ particles from K to K* system they are transformed from the K* frame to the $K_0$ frame according to formulas

$$p_{xi}^0 = -p_{xi}^* \cos\theta \cos\varphi - p_{yi}^* \sin\varphi - p_{zi}^* \sin\theta \cos\varphi, \qquad (6)$$

$$p_{zi}^0 = p_{xi}^* \sin\theta - p_{zi}^* \cos\theta, \qquad (7)$$

$$p_{yi}^0 = -p_{xi}^* \cos\theta \sin\varphi + p_{yi}^* \cos\varphi - p_{zi}^* \sin\theta \sin\varphi, \qquad (8)$$

where $p_{xi}^*, p_{yi}^*, p_{zi}^*$ are momentum projections of i-th $\alpha$-particle (in the K* frame), and $\theta$, $\varphi$– angles are determined as

$$\cos\theta' = -P_z / \sqrt{P_x^2 + P_y^2 + P_z^2}, \qquad (9)$$

$$\cos\varphi' = -P_x / \sqrt{P_x^2 + P_y^2}, \qquad (10)$$

$$\sin\varphi' = -P_y / \sqrt{P_x^2 + P_y^2}, \qquad (11)$$

where $P_x$, $P_y$, $P_z$ momentum projections of $^{12}C^*$ nucleus fragment.

## 1.2 The decay: $^{12}C^*$ nucleus fragment $\to Be^4 + \alpha \to 3\alpha$

The process (4) of decay of $^{12}C^*$ nucleus fragment in to the excited nucleus with quantum numbers of beryllium $Be^4$ and $\alpha_1$-particle is generated with probability 0.6. In this decay the production angles of $\alpha_1$-particle $\theta'_1$ and $\varphi'_1$ in the $^{12}C^*$ nucleus fragment rest frame are generated in accordance with isotropic angular distribution. The z' axis of $^{12}C^*$ nucleus fragment rest frame is parallel to its momentum in the $K_0$ frame and y' axis is directed along to the vector product $\vec{z}^* \times \vec{z}_0$. The projections of $\alpha_1$ and $Be^4$ momenta are transformed from the $^{12}C^*$ nucleus fragment rest frame into $K_0$ rest frame. The same way is used to generation of decay of $Be^4$ into two $\alpha$-particles.

# Captions

Fig.1 The dependence of an average value of an α-particles transverse momentum on the excitation energy $\Delta E^*$.

Fig.2 The dependence of an average absolute value of an α-particles longitudinal momentum in the system of a zero longitudinal momentum of a $^{12}C^*$ nucleus fragment on the excitation energy $\Delta E^*$.

Fig.3 The distribution on the azimuthal angle between α-particles pairs at the excitation energy $\Delta E^* < 15$.

Fig.4 The distribution on the azimuthal angle between α-particles pairs at the excitation energy $\Delta E^* > 15$.

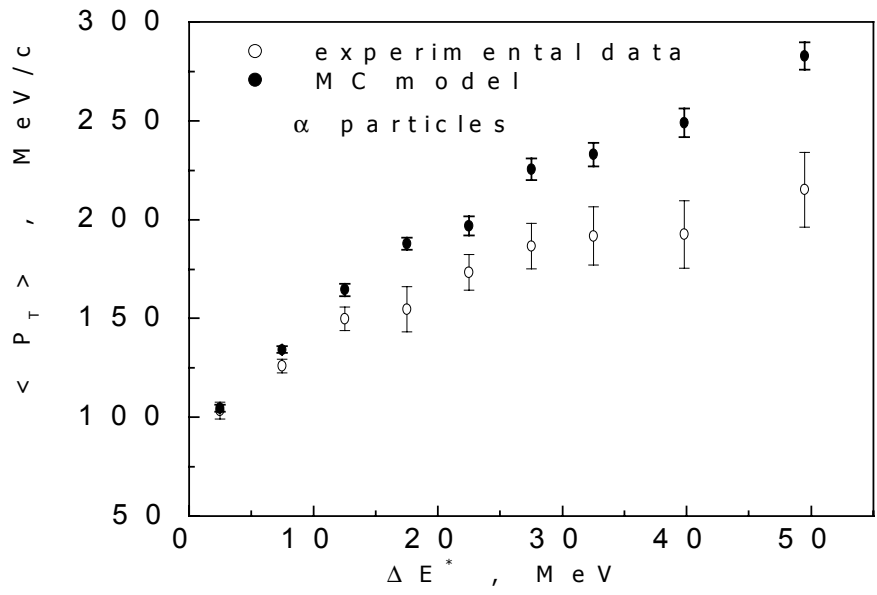

Fig.1

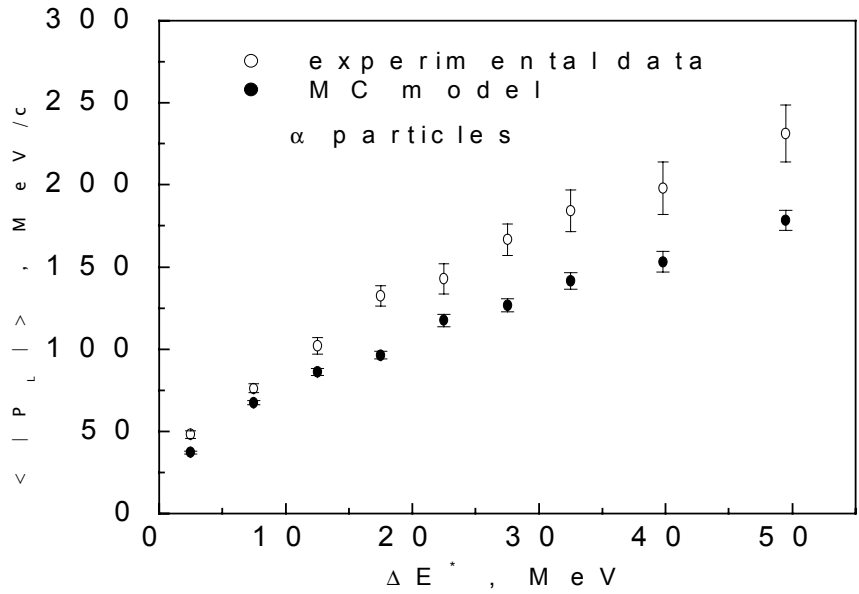

Fig.2

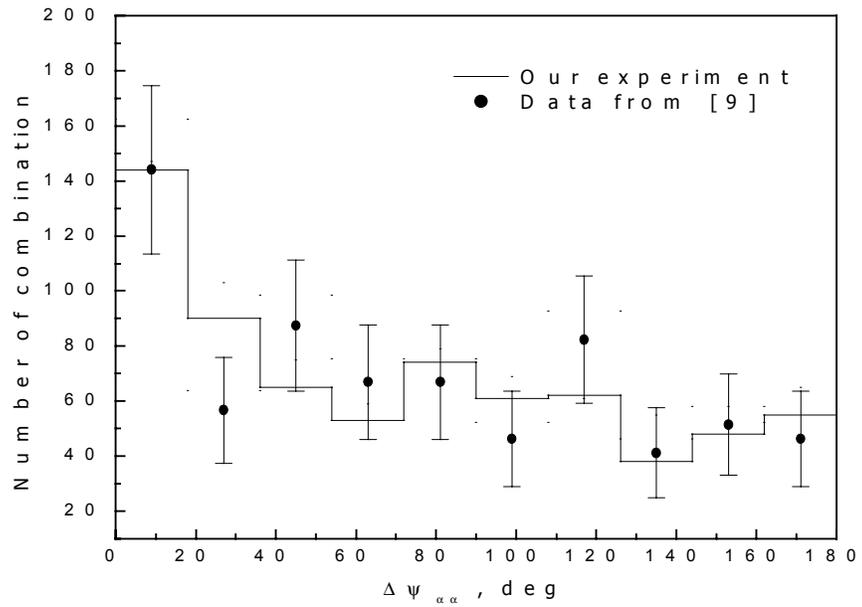

Fig.3

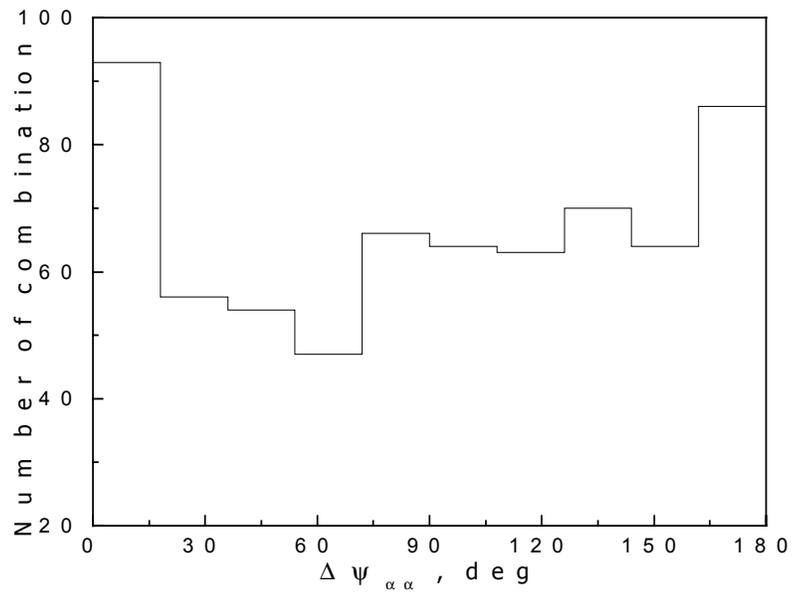

Fig.4